\documentclass[prd,onecolumn,nofootinbib,showpacs]{revtex4}
\usepackage{epsfig,graphicx,bm}
\usepackage[latin2]{inputenc}
\usepackage{color}
\usepackage{t1enc}
\usepackage{amssymb}
\usepackage{amsmath}
\usepackage{graphicx}
\usepackage{bm}

\newcommand{\bea}{\begin{eqnarray}}
\newcommand{\eea}{\end{eqnarray}}
\newcommand{\be}{\begin{equation}}
\newcommand{\ee}{\end{equation}}

\newcommand{\slp}{\mbox{\,\slash \hspace{-0.5em}$p$}}

\newcommand{\p}{{\bf p}}
\newcommand{\intsum}{\sum\nolimits_{\p} \hspace{-1.9em}\int \hspace{0.6em}}
\def\lag{\langle}
\def\rag{\rangle}
\newcommand{\no}{{\nonumber}}
\newcommand{\Tr}{\mathop{\mathrm{Tr}}}

\newcommand{\muI}{\mu_{\text{I}}}
\newcommand{\muB}{\mu_{\text{B}}}
\newcommand{\muIB}{\mu_{\text{I},\text{B}}}
\newcommand{\im}{\mathrm{i}}

\newcommand{\rt}[1]{{}}

\begin{document}
\allowdisplaybreaks

\title{Pion condensation in the two--flavor chiral quark model at
    finite baryochemical potential}

\author{T. Herpay}
\email{herpay@cleopatra.elte.hu}
\author{P. Kov\'acs}
\email{kpeti@cleopatra.elte.hu}
\affiliation{Research Institute for Particle and Nuclear Physics
of the Hungarian Academy of Sciences, H-1525 Budapest, Hungary}
\affiliation{Research Group for Statistical and Biological Physics 
of the Hungarian Academy of Sciences,
H-1117 Budapest, Hungary}

\begin{abstract}
Pion condensation is studied at one--loop level and nonzero
baryochemical potential in the framework of two flavor constituent
quark model using the one--loop level optimized perturbation theory
for the resummation of the perturbative series. A Landau type of
analysis is presented for the investigation of the phase boundary
between the pion condensed/non-condensed phases. The statement 
that the condensation starts at $\muI = m_{\pi}$ is slightly
modified by one--loop corrections. The second order critical surface
is determined and analysed in the $\muI-\muB-T$ space. The $\muI$
dependence of the one--loop level charged pion pole masses is
also studied.
\end{abstract}


\maketitle

\section{INTRODUCTION}

From the early 70's, when the idea of pion condensation was first
suggested \cite{migdal71}, many investigations have been performed
in the field. Right in the early works, the possibility emerged that pion
condensation might have a significance in the evolution of compact
stars \cite{sawyer72,scalapino72,migdal73}. Beside compact stars,
pion condensation can occur in asymmetric nuclear matter as well as in
heavy ion collisions at intermediate energies.
While the phases of QCD can be well described by perturbation theory
at extreme high temperatures and densities, at moderate temperatures and
densities a perturbative description is not possible,
here other methods like lattice field theory and effective field
theories can be applied. In lattice field theory a second order phase
transition was found from normal phase to pion condensed phase in two
flavor QCD at a critical isospin chemical potential, which is about
the pion mass \cite{kogut02a,kogut02b,kogut04}. Pion and
in some cases kaon condensation was also investigated in numerous
effective models such as chiral perturbation theory \cite{son01,
  kogut01}, ladder QCD \cite{barducci03}, random matrix method
\cite{klein03}, NJL model (in the mean field approximation)
\cite{barducci04,barducci05,ebert06a,ebert06b}, PNJL model
\cite{zhang06} and linear sigma model \cite{he05, andersen06,
  mao06}. These investigations mostly focus on properties of
asymmetric hadronic matter, however if one wants to describe for
instance a neutron star, which is electrically neutral
on average, then the condition of charge neutrality as well as
$\beta$--equilibrium must be imposed explicitly as discussed in
\cite{ebert06a,andersen07}.

Our main goal here is to go beyond the present status of the field
 and determine the phase boundary between 
the pion condensed and non--condensed phases in the $\muI-\muB-T$
space at one--loop level and to calculate the $\muI$ dependence of the
charged pion masses beyond tree level. This work is the continuation 
of our previous works \cite{kovacs06,kovacs07}, 
in the sense that we use the same
framework of constituent quark model and the optimized perturbation
theory (OPT) \cite{chiku98} as a resummation technique, which preserves
the  perturbative renormalizability as well as the symmetries of the
model. However there we used three flavor model, while the
present investigation deals with an effective model of 2--flavor
QCD. This approximation is sufficient to see the main properties of
pion condensation.

 The paper is organized as follows. In Section II,
we introduce the model and discuss its renormalization in detail in
case when not only scalar but also pseudoscalar (pion) condensate is
present. We treat bosons at one--loop level, while the constituent
quarks are considered at tree level. In Section III, we present the one--loop
equations, which describe our system, namely the equations of state
for the scalar and pion condensates and the equation for the resummed
pion mass. Moreover we show that a Ward identity holds, from which the
Goldstone theorem also follows. Then we expand our equations
in powers of the pion condensate, with which we restrict ourselves to
a Landau--type analysis of the phase transition. At the end of this
section a simple formula for the boundary of the pion condensation
domain is established. In Section IV we diagonalize the mixed boson
and fermion propagators, which are determined also for small values of
the pion condensate. Section V is dedicated to the parameterization
of the model, i.e. we fix the parameters of the Lagrangian at zero
temperature and vanishing chemical potentials at one--loop level. Here
we also discuss the choice of the renormalization scales. In Section
VI we present the numerical results and finally we conclude in Section
VII.

\section{THE MODEL AND ITS RENORMALIZATION}
\label{model}

Our starting point is the renormalized $SU(2)_{L}\times SU(2)_{R}$
symmetric Lagrangian with explicit symmetry breaking term, which
contains meson as well as quark fields \cite{levy67}. Without the
symmetry breaking term the symmetry group of
the light meson sector ($\pi,\,\sigma$) is  $O(4)$, and the Lagrangian is 
\bea
\label{lag}
{\cal L}&=&\frac{1}{2}\left(\partial_\mu \phi \partial^\mu
\phi-m^2\phi^2\right)-\frac{\lambda}{4}\phi^4+h\phi_0+ \im\bar{\psi}
\gamma_\mu \partial^\mu
\psi-\frac{g_F}{2}\bar{\psi}T_{\alpha}\phi_{\alpha}\psi \no\\
&+&\frac{1}{2}\left(\delta Z
\partial_\mu \phi
\partial^\mu\phi-\delta
  m^2\phi^2\right)-\frac{\delta\lambda}{4}\phi^4,
\eea
where $\psi=(u,\,d)^T$ are the $SU(2)$ doublet quark fields,
$\phi=(\phi_0,\phi_1, \phi_2, \phi_3)\equiv(\sigma, \pi_1, \pi_2,
\pi_3)$ are the scalar sigma  and  pseudoscalar pion fields, $h$ is
the symmetry breaking external field, and $T_{\alpha}=(\tau_0,
\im{\bm\tau}\gamma_5), \quad \alpha=0\dots 3$ are the quark--boson
coupling matrices. In the Lagrangian above the couplings $m^2$,
$\lambda$  and the $\phi$ field are renormalized (finite) quantities,
while $\delta m^2, \delta\lambda$, and $\delta Z$ are
the corresponding (infinite) counterterms. Since we treat
quarks only at tree level no wave function renormalization constant is
necessary for the fermionic fields.

We use the grand canonical
generating functional of the renormalized Lagrangian in order to
introduce the baryon and isospin chemical potentials as well as the
scalar and pion condensates. As one shall see, one consequence of the
presence of pion condensation amounts to the appearance of the
non--trivial wave function renormalization constant in the equation of
state of the condensate. At finite temperature and densities the grand
canonical generating functional, which determines the $N$--point
functions of the fields, is 
\be
\label{gen_func}
{\cal Z}=\int {\cal D}\phi {\cal D}\Pi  {\cal D}{\bar{\psi}} {\cal
  D}\psi\,\exp\left[{\im\int_0^{-\im\beta} dt \int
  d^3x(\Pi\dot{\phi}+\im\bar{\psi}\gamma_0\dot{\psi} - {\cal H}+\muB
  Q_{\mathrm{B}}+\muI Q_{\mathrm{I}})}\right] ,
\ee
where the Hamiltonian of the system is
\bea
{\cal H} &=&
\frac{1}{2}\left(\Pi^2+(\nabla\phi)^2+m^2\phi^2\right)+\frac{\lambda}{4}\phi^4-h\phi_0
+\im\bar{\psi}\gamma_i\partial_i\psi+\frac{g_F}{2}\bar{\psi}T_{\alpha}\phi_{\alpha}\psi \nonumber\\
&-&\frac{1}{2}\delta Z\Pi^2+\frac{1}{2}\delta Z(\nabla\phi)^2 +\frac{\delta \lambda}{4}\phi^4+
\frac{1}{2} \delta m^2 \phi^2,\quad\quad \text{\footnotesize $(i=1,2,3)$}
\eea
and the canonical momenta of the scalar fields are
defined by
\be
\Pi=\frac{\delta {\cal L}}{\delta \dot{\phi}}=(1+\delta Z)\dot{\phi}.
\ee

In (\ref{gen_func}) $Q_B$, $Q_I$ are the conserved baryon and isospin charges defined by
the zeroth component of the corresponding Noether currents of the
symmetric Lagrangian
 \bea
Q_{B}&=&\int\mathrm{d}^3
x \frac{1}{3}(u^\dag u +{d}^\dag d), \\
Q_{I}&=&\int\mathrm{d}^3
x\left[(1+\delta Z)(\pi_1\dot\pi_2-\pi_2\dot\pi_1)+\frac{1}{2}\left(u^\dag
  u-d^\dag d\right)\right].
\eea
At small temperatures, when $h\ne0$ or $h=0$ and $m^2<0$ the original
$O(4)$ symmetry of the meson sector breaks down to $O(3)$, which means that
the $\sigma$ scalar field has a nonzero vacuum expectation value
(vev), thus $\lag \phi_0 \rag  \equiv \lag \sigma \rag \equiv v \ne
0$. Moreover, if the isospin chemical potential $\muI$ is sufficiently
large, charged pions condensate (Bose-condensation) and another
nonzero vacuum expectation value occurs, which additionally breaks
down the symmetry to $O(2)$. If one initially introduces vev to all
the four bosonic fields it can be shown that because of the remaining
$O(2)$ symmetry the fields can be transformed in such a way that
only two nonzero expectation value remains, from which one  is
the aforementioned $v$, and the other one can be either $\lag \phi_1 \rag$ or
$\lag \phi_2 \rag$, i.e. one can entirely transform out $\lag \phi_3
\rag$. Hence, we can choose $\lag \phi_1 \rag \equiv \lag \pi_1 \rag
\equiv \rho \ne 0$ and $\lag \phi_i \rag \equiv \lag \pi_i \rag = 0$
for $i=2,3$ (see for e.g.\cite{andersen06}).

Shifting the fields by their expectation values in Eq.\eqref{gen_func}
the generating functional can be written as
\be
\label{shift_Z}
{\cal Z}=\int {\cal D}\phi {\cal D}{\bar{\psi}} {\cal
  D}\psi\left[\mathrm{e}^{-\int_0^{-\im\beta} dt \int
  d^3x \bar{\psi}{G^{f}}^{-1}\psi} \mathrm{e}^{ -\im\int_0^{-\im\beta} dt \int d^3x
  \tilde{{\cal L}}_I }\int {\cal D} \Pi \mathrm{e}^{i\int_0^{-\im\beta} dt \int
  d^3x( \Pi\dot{\phi} - \tilde{{\cal
      H}}_B)}\right] .
\ee
Here ${G^{f}}^{-1}$ is the tree level fermion propagator matrix, which
after going to Fourier space and introducing the
$\omega_n=(2n+1)\beta$ fermionic Matsubara frequencies, is given as
(see for e.g.~\cite{he05}),
\be
\label{prop_f}
\im{G_{ij}^{f}}^{-1} =\begin{pmatrix}
(-\im\omega_n+\frac{1}{3}\muB+\frac{1}{2}\muI)\gamma_0-\gamma_i
p_i-\frac{g_{F}}{2}v & -\im\frac{g_{F}}{2}\gamma_5\rho
\\ -\im\frac{g_{F}}{2}\gamma_5\rho & (-\im\omega_n+\frac{1}{3}\muB-
\frac{1}{2} \muI)\gamma_0-\gamma_i
p_i-\frac{g_{F}}{2}v \end{pmatrix},
\ee
where $p_i$ is the spatial momentum. In \eqref{shift_Z} $\tilde{{\cal
    H}}_B = \tilde{{\cal H}}_B^{\Pi} + \tilde{{\cal H}}_{B,2}$
contains all terms depending on the canonical momenta
and bosonic fields up to second order. Other terms are
contained in the  $\tilde{{\cal L}}_I$. The non--zero condensates
generate three point couplings, which can be directly read off from
the explicit form of $\tilde{{\cal L}}_I$ \eqref{e:sh2}. The
combination
$\Pi\dot{\phi}-\tilde{{\cal H}}_B^{\Pi}$ can be written as
\bea
\label{gen_mom_part}
\Pi\dot{\phi}-\tilde{{\cal H}}_B
=&-&\frac{1}{2}(1-\delta Z)\left(\Pi_0^2-2\Pi_0(1+\delta Z)\dot{\phi}_0\right)\no\\
&-&\frac{1}{2}(1-\delta Z)\left(\Pi_3^2-2\Pi_3(1+\delta Z)\dot{\phi}_3\right)\no\\
&-&\frac{1}{2}(1-\delta Z)\left(\Pi_1^2-2\Pi_1(1+\delta
  Z)(\dot{\phi}_1-\muI\phi_2)\right)\no\\
&-&\frac{1}{2}(1-\delta Z)\left(\Pi_2^2-2\Pi_2(1+\delta
  Z)(\dot{\phi_2}+\muI(\phi_1+\rho))\right).
\eea
After completing into whole squares and performing the
integration over the canonical momenta in \eqref{gen_mom_part}, one
can identify the tree level bosonic inverse propagator matrix and from
the linear terms the tree level equations of states (EoS).

 The $4\times 4$ inverse boson propagator (${G^{b}}^{-1}$) in Fourier space
with the introduction of the $\omega_n=2\pi n\beta$ bosonic Matsubara
frequencies splits up into the inverse propagator of $\pi_3$ and a
$3\times 3$ coupled inverse propagator matrix of the remaining three
bosonic fields $\pi_1, \pi_2, \sigma$. In the $\pi_{1,2}$ sector we
switch to the charged $\pi^{\pm} = (\pi_1\mp \im\pi_2)/\sqrt{2}$ base
and use the following relabeling $i,j=1,2,3,4 \longrightarrow
\pi^+,\pi^-,\sigma,\pi_3$. In this way the non zero elements of
${G_{ij}^{b}}^{-1}$ are
\bea
\label{prop_b}
\im{G_{44}^{b}}^{-1} &=& (-\im\omega_n)^2-E_{\pi_3}^2,\\
\im{G_{kl}^{b}}^{-1}&=&\begin{pmatrix}
(-\im\omega_n-\muI)^2-E_{\pi_3}^2-\lambda\rho^2 &
-\lambda\rho^2 & -\sqrt{2}\lambda v\rho \\
-\lambda\rho^2 & (-\im\omega_n+\muI)^2-E_{\pi_3}^2-\lambda\rho^2
& -\sqrt{2}\lambda v\rho \\
-\sqrt{2}\lambda v\rho & -\sqrt{2}\lambda v\rho &
(-\im\omega_n)^2-E_{\pi_3}^2-2\lambda v^2
\end{pmatrix}\no,
\eea
where $k,l=1,2,3$, $E_{\pi_3}=\sqrt{\p^2+m_{\pi_3}^2}$, and the
tree level $\pi_3$ mass square is $m_{\pi_3}^2 = m^2 + \lambda(v^2 +
\rho^2)$.

As mentioned earlier in \eqref{gen_mom_part} the coefficients
of the terms linear in the corresponding fields $\phi_0$ ($\sigma$)
and $\phi_1$ ($\pi_1$) determine the two non--trivial equations of
states:
\bea
  {\textrm{EoS}}_{\sigma}^{\text{tree}}&=& v(m^2+\lambda(v^2+\rho^2))-h=0,\\
  {\textrm{EoS}}_{\pi_{1}}^{\text{tree}}&=& \rho(m^2+\lambda(v^2+\rho^2)-\muI^2(1+\delta Z))=0.
\eea
In the second EoS the wave function renormalization constant $\delta
Z$ is explicitly written ---which unavoidably occurs when field
renormalization is introduced followed by the canonical way of
introducing chemical potentials--- even if it is a higher
order term, because we would like to emphasize the importance of its
presence. From the one--loop fermionic contribution of
the second EoS, as it will be seen later, a $\muI^2$ proportional
divergence arise, which has to be absorbed into some infinite
counterterm, and this counterterm will be the $\delta Z$
renormalization constant. In other words the lack of $\delta Z$
would result in an uncanceled divergence in the second EoS.

The infinite parts of the counterterms can be obtained  by
requiring the finiteness of the perturbative $N$--point functions (in our
case the propagator and the four point boson vertex) in
the symmetric phase ($v=\rho=0$) at $T=\muB=\muI=0$. Practically it
is easier to obtain the infinite parts from the one--loop EoS
(see Sec. \ref{eqns_1_loop}). Accordingly, the following counterterms
can be found up to one--loop order with cutoff regularization,
\bea
  \delta m^2
  &=&-6\lambda(\Lambda^2-m^2\ln\frac{\Lambda^2}{l_b^2}) +
  \frac{g_F^2}{4\pi^2} N_c \Lambda^2 ,\label{e:cnt}\no\\
  \delta \lambda &=& 12\lambda^2\ln\frac{\Lambda^2}{l_b^2}-
  \frac{g_F^2}{32\pi^2} N_c\ln\frac{\Lambda^2}{\mathrm{e}l_f^2},\\
  \delta Z &=& -N_c\frac{g_F^2}{16\pi^2}\ln\frac{\Lambda^2}{l_f^2\mathrm{e}^2},\no
\eea
where $\Lambda$ is the three dimensional cutoff in momentum space,
while $l_b$ and $l_f$ are the bosonic and fermionic renormalization
scales, respectively. These counterterms cancel by definition all
divergences at one--loop level and at $v=\rho=T=\muB=\muI=0$ and in
the broken phase the same counterterms can be used with a slight
modification due to the necessary resummation (see
Sec. \ref{eqns_1_loop}). The finite parts of $\delta
m^2$, $\delta \lambda$ and $\delta Z$ are determined by the
parameterization of the model in the broken phase (see Sec. \ref{diag}).

\section{EQUATIONS AT ONE--LOOP LEVEL}
\label{eqns_1_loop}

As it is well known from finite temperature field theory, tree level
mass squares can become negative in the broken phase as the
temperature increases. Accordingly some sort of resummation
of the perturbative propagator is needed (see for example in
\cite{dolan74}). We perform this using the optimized perturbation
theory \cite{chiku98}, where a temperature (and chemical potential)
dependent mass term is introduced in the Lagrangian and the difference
between the original and the new mass parameter is treated as a
(finite) higher order counterterm,
\be
\label{lag2}
{\cal L}_{mass}=-\frac{1}{2}m^2\phi^2=-\frac{1}{2}M^2(T,\mu)\phi^2 -
\frac{1}{2}\Delta m^2 (T,\mu)\phi^2.
\ee
Here $\Delta m^2$ is the finite ``one--loop level'' counterterm. The new
mass parameter is determined by requiring that
the inverse one--loop level $\pi_3$ propagator at zero external
momentum stays equal to its tree level value (fastest apparent convergence FAC)
\be
M^2_{\pi_3}\equiv M^2 + \delta m^2 + (\lambda +
\delta\lambda)(v^2+\rho^2) + \Sigma_{\pi_3}(\omega=\p=0, M^2,T,\mu) +
\Delta m^2 = M^2 + \lambda(v^2+\rho^2)\equiv m_{\pi_3}^2, \label{e:sum1}
\ee
where we indicated the resummed mass dependence of the
self energy ($\Sigma_{\pi_3}$). Strictly speaking
the OPT at one--loop order replaces $m$ with $M(T,\mu)$ also in the
internal tree level propagator lines. Since $\pi_3$ is not a mixed
state, $M(T,\mu)$ can be expressed through the tree
level $\pi_3$ mass, which equals its one--loop level value by the
condition \eqref{e:sum1}, and thus the diagonal part of the boson
propagator \eqref{prop_b} can be written as a function of
the $\pi_3$ mass instead of $M(T,\mu)$,
\be
\begin{array}{lll}
\im{G_{11}^{b}}^{-1}=(-\im\omega_n-\muI)^2 - \p^2 -
  m_{\pi_3}^2 - \lambda\rho^2, & & i{G_{22}^{b}}^{-1}=(-\im\omega_n +
  \muI)^2-\p^2 - m_{\pi_3}^2 - \lambda\rho^2 \\
\im{G_{33}^{b}}^{-1}=(-\im\omega_n)^2 - \p^2-m_{\pi_3}^2-2\lambda v^2, & &
\im{G_{44}^{b}}^{-1} =(-\im\omega_n)^2 - \p^2 - m_{\pi_3}^2. \end{array}\label{e:prop_b2}
\ee
Here and in the following the $\pi_3$ mass is denoted by $m_{\pi_3}$
due the PMS relation ($m_{\pi_3}=M_{\pi_3}$). In this way $M(T,\mu)$
is eliminated from \eqref{e:sum1} and the resummed $\pi_3$ mass is
determined by the equation
\be
\label{e:mass_eq}
m^2_{\pi_3}(T,\mu) = m^2 + \delta m^2 + (\lambda + \delta\lambda)
(v^2+\rho^2) + \Sigma_{\pi_3}(\omega=\p=0,m_{\pi_3}^2,T,\mu),
\ee
where $\Sigma_{\pi_3}$ now depends on $m_{\pi_3}$ through the
 ``resummed tree level'' propagator matrix \eqref{e:prop_b2}. This
means that the above procedure makes the $\pi_3$ propagator
($G_{44}^{b}$) selfconsistent at $p=0$, while the mixed sector of the
boson propagator is partially resummed due to its $m_{\pi_3}$
dependence.

The self energy in (\ref{e:mass_eq}) contains bosonic as
well as fermionic loop integrals,
\bea
\Sigma_{\pi_3}(\omega=\p=0,m_{\pi_3}^2,T,\mu)& = &\lambda + \lambda
\intsum \Tr \{T^b G^b(\omega_n,\p,\muI)\}\no\\
& + & \lambda^2\intsum B^b_{ij} B^b_{kl} \left( G^b_{ik}(\omega_n,
\p, \muI) G^b_{jl}(\omega_n, \p,\muI) + G^b_{il}(\omega_n, \p, \muI)
G^b_{jk}(\omega_n,\p,\muI)\right) \no \\ 
& + & g_F^2 \intsum G^f_{il}(\omega_n, \p, \muI,
\muB) B^f_{ij} G^f_{jk}(\omega_n, \p, \muI, \muB) B^f_{kl},\label{e:sum2}
\eea
where $T^b$, $B^{b,f}$ denotes the coupling matrices which arise from
\eqref{gen_mom_part} and are listed in the appendix.
It is worth to note that the divergences of the self energy are
cancelled by the infinite perturbative counterterms if one replaces
$m$ with $M(T,\mu)$ in \eqref{e:cnt}. In this case, it seems that the
counterterms are temperature and/or chemical potential dependent,
however it can be proved order by order that all $T$ and/or $\mu$
dependent infinities are cancelled by higher order contributions (see
for e.g. in \cite{chiku98} and \cite{jakovac04}).

The scalar
condensate $v$ is determined by the vanishing of the one--loop
level one point function of $\sigma$ (EoS$_\sigma$ at
one--loop level),
\be
v\left(m^2 + \delta m^2 + (\lambda+\delta\lambda) (v^2 + \rho^2) +
  \lambda \intsum \Tr\{H^b
  G^b(\omega_n, \p, \muI)\} + g_F\intsum \Tr \{H^f
  G^f(\omega_n, \p, \muI, \muB)\}\right) = h, \label{e:eos2}
\ee
where $H^{b,f}$ can be found in the appendix. Comparing
\eqref{e:sum2} and \eqref{e:eos2} one can recognize a Ward identity
which  connects  the symmetry breaking external field
with the propagator of $\pi_3$ at zero external momentum,
\be
v m^2_{\pi_3}=h.\label{e:ward}
\ee
This relation is a consequence of the remaining $O(2)$ symmetry, which
is an axial vector rotation around the third isospin axis from the point
of view of the chiral symmetry. Moreover \eqref{e:ward} guarantees the
Goldstone theorem for this degree of freedom (neutral pion).

The pion condensate is determined through the $\pi_1$ one point
function, which is the EoS$_{\pi_1}$ at one--loop level, 
\bea
&&\rho\left(m^2 + \delta m^2 + (\lambda + \delta\lambda)(v^2 + \rho^2) -
  \muI^2(1+\delta Z) + \lambda \intsum \Tr\{R^b
  G^b(\omega_n, \p, \muI)\}\right.\no\\
&&\left.+g_F \intsum \Tr\{R^f G^f(\omega_n, \p, \muI,
  \muB)\}\right)=0, \label{e:eosr}
\eea
where and $R^{b,f}$ can be found in the appendix. Here the infinite part of
$\delta Z$ just cancels the $\muI$ dependent divergence of the fermion
loop integral.

The calculation of the loop integrals in \eqref{e:sum2} and
\eqref{e:eosr} requires the diagonalization of the propagators and the
corresponding transformation of the coupling matrices due to the
non--diagonal matrix elements in \eqref{prop_b} and
\eqref{prop_f}. The diagonalization itself is a straightforward
calculation, however the eigenvalues of the boson propagator
\eqref{prop_b} are non--rational functions of $\omega_n$. Thus
performing the Matsubara sums is a very complicated task and it is
beyond the scope of our paper. In order to avoid these difficulties the
diagonalization was accomplished only for small $\rho$. As one shall
see in the next section this method leads to ordinary Matsubara
frequency dependence in the propagators. For that very reason we
restricted ourselves to a Landau--type analysis of the pion condensation
in a small vicinity of the phase boundary. Up to second order in $\rho$
the mass equation \eqref{e:mass_eq} can formally be written as
\be
\label{e:pi3_pert}
m^2_{\pi_3}=m^2 + \lambda v^2+t^{(0)}(m^2_{\pi_3},v,T,\muIB) +
(\lambda+t^{(2)}(m^2_{\pi_3}, v, T, \muIB))\rho^2,
\ee
while  \eqref{e:eos2} goes over into the simple form of
\eqref{e:ward} due to the Ward identity. At the same order in $\rho$
 \eqref{e:eosr} can be rewritten as, 
\be
\label{e:eosr2}
\rho\left[\mu_I^2- m^2- \lambda v^2 -r^{(0)}(m^2_{\pi_3}, v, T, \muIB)
  -(\lambda+r^{(2)}(m^2_{\pi_3},v,T,\muIB))\rho^2+{\cal O}(\rho^4)\right]=0
\ee
By virtue of the above equation the pion condensate may have non--zero
value only if the roots of the expression in the square bracket are
real. Assuming that $\lambda+r^{(2)}>0$, Eq.\eqref{e:eosr2} yields 
\be
\label{e:rho_pert}
\rho=\sqrt{\frac{\mu_I^2 -m^2- \lambda v^2 - r^{(0)}(m^2_{\pi_3}, v, T,
    \muIB)} {\lambda + r^{(2)}(m^2_{\pi_3},v, T, \muIB)}} 
\ee
if $\mu_I^2 -m^2-\lambda v^2-r^{(0)} > 0$. This means that in this
case the transition is of second order supposing that the coefficient
of the fourth order term in \eqref{e:eosr2} is negative. Moreover if
$\lambda +r^{(2)}<0$ and $\mu_I^2 -m^2 -\lambda v^2-r^{(0)} < 0$
(keeping that the fourth order term is negative) it can be seen that the
equation can have two nonzero roots, which suggests
first order phase transition. For the calculation of the $t^{(0,2)}$
and $r^{(0,2)}$ coefficients in \eqref{e:pi3_pert} and
\eqref{e:rho_pert} diagonalization of the boson and fermion propagator
up to order $\rho^2$ is needed, which will be presented in the next
section.

\section{DIAGONALIZED PROPAGATORS FOR SMALL $\mathbf{\rho}$}
\label{diag}

In our approach, as was discussed previously, the next step
is to determine the eigenvalues of the propagator matrices for small
$\rho$ values, in other words to find the propagating eigenmodes
with help of some suitable linear transformation of the original
fields perturbatively in $\rho$. This step is not necessary if one would like to
calculate the effective potential (see for e.g. \cite{he05}). In case
of the bosonic propagator the transformation matrix up to
${\cal O}(\rho^3)$ is found to be,
\be
\label{trans_b}
O_{B}=\begin{pmatrix}
1-\left|a\right|^2\rho^2 & b(1-2av)\rho^2 & -\sqrt{2}a\rho \\
b^{*}(1-2a^{*}v)\rho^2 & 1-\left|a\right|^2\rho^2 & -\sqrt{2}a^{*}\rho \\
\sqrt{2}a\rho & \sqrt{2}a^{*}\rho & 1-2\left|a\right|^2\rho^2
\end{pmatrix}+{\cal O}(\rho^3),
\ee
where $a=a(\omega_n,\mu)=\lambda v/(\mu^2+2\lambda v^2+2\im\omega_n\mu)$
and $b=b(\omega_n,\mu)=\im\lambda/(4\mu\omega_n)$. As it can be checked
$O_B$ is not a unitary transformation and it is important to note that
$O_B$  depends on the Matsubara frequency $\omega_n$. With this
transformation
$O_B\cdot (\im {G_{\pi^{+},\pi^{-},\sigma}^{b}}^{-1}) \cdot O_B^{-1}=
\mathrm{diag}(\im\tilde{G}_{\pi^+}^{-1}, \im\tilde{G}_{\pi^-}^{-1},
\im\tilde{G}_{\sigma}^{-1}) + {\cal O}(\rho^3)$, where the tilde reminds
us that these propagators belong to the transformed (propagating)
particles. It is worth to note that the new $\pi^+$ and $\pi^-$
particles are no longer charge conjugates of each other, which is a
natural consequence of the presence of the pion
condensate. After calculating the inverses perturbatively, the transformed
$\pi^+$, $\pi^-$ and $\sigma$ bosonic propagators are given by
\bea
  \im\tilde{G}_{\pi^+}&=&\frac{1}{(\omega_n+\im\muI)^2+E_{\pi}^2}-
  \rho^2\frac{\lambda(2\muI^2+2\lambda v^2 -
    4\im\muI\omega_n)}{((\omega_n+\im\muI)^2+E_{\pi}^2)^2(\muI^2 +
    2\lambda v^2 - 2\im\muI\omega_n)}+{\cal O}(\rho^4),\\
  \im\tilde{G}_{\pi^-}&=&\frac{1}{(\omega_n-\im\muI)^2+E_{\pi}^2} -
  \rho^2\frac{\lambda(2\muI^2+2\lambda v^2 + 4\im\muI\omega_n)}
      {((\omega_n-\im\muI)^2+ E_{\pi}^2)^2(\muI^2+2\lambda v^2 +
        2\im\muI \omega_n)}+{\cal O}(\rho^4),\\
  \im\tilde{G}_{\sigma}&=&\frac{1}{\omega_n^2+E_{\sigma}^2}-\rho^2\frac{\lambda(\muI^2
      + 2\lambda v^2) (\muI^2 + 6\lambda v^2 + 4\muI^2\omega_n^2)}
    {(\omega_n^2 +E_{\sigma}^2)^2 ((\muI^2+2\lambda v^2)^2 +
      4\muI^2\omega_n^2)}+{\cal O}(\rho^4),
\eea
while the $\pi_3$ propagator is
\be
  \im G_{\pi_3}=\frac{1}{\omega_n^2+E_{\pi}^2}-\rho^2\frac{\lambda}
    {(\omega_n^2 +E_{\pi}^2)^2}+{\cal O}(\rho^4).
\ee
In case of the fermionic inverse propagator matrix the diagonalization
must be performed cautiously due to the presence of the non
commuting Dirac matrices. The clearest approach is to solve the equation
$O_F(\im G_F^{-1})O_F^{-1}=\mathrm{diag}$ for $O_F$ directly. In this way
$O_F$ is found to be
\be
\label{trans_f}
  O_{F}=
\begin{pmatrix}
  1+\frac{g_F^2}{32k_0^2}\rho^2 & -\im\frac{g_F}{4k_0}\gamma_0\gamma_5\rho\\
  -\im\frac{g_F}{4k_0}\gamma_0\gamma_5\rho & 1+\frac{g_F^2}{32k_0^2}\rho^2
\end{pmatrix},
\ee
where $k_0=(-\im\omega_n+\frac{1}{3}\muB)\gamma_0$ and the matrix is hermitian.
After performing the inverse perturbatively the fermionic propagators
are given by
\be
 \im\tilde{G}_{u/d}=-\frac{1}{\slp_{u/d}-m_f}- \rho^2 \frac{g_F^2}{8k_0}
  \frac{1}{\slp_{u/d}-m_f} \gamma_0\frac{1}{\slp_{u/d}-m_f}
\ee
where $\slp_{u/d}=(-\im\omega_n+\mu_{u/d})\gamma_0-\gamma_ip_i$ and
$\mu_{u/d}=\muB/3\pm\muI/2$.

In the appendix it is shown that all integrandus appears in the
one--loop equations \eqref{e:sum2}, \eqref{e:eos2} and \eqref{e:eosr}
can be written as traces over flavor space (in
case of fermions traces also concern Dirac indices). In
this way one can insert the bosonic/fermionic
 transformation matrices given in \eqref{trans_b} and \eqref{trans_f}
under the traces and transform
the propagators within into diagonal form, which will lead to
transformation of the corresponding coupling matrices. Thus for
instance the trace in \eqref{a:bubble_tr} can be written as
\be
 \Tr \{{B^b}^{\prime} G^b\} = \Tr\{{B^b}^{\prime} O_B^{-1} O_B G^b
 O_B^{-1} O_B \} = \Tr\{O_B{B^b}^{\prime}O_B^{-1}\tilde{G}^b \} = \Tr
 \{\tilde{B}^b \tilde{G}^b \}, 
\ee
where $\tilde{G}^b = \mathrm{diag}(\tilde{G}_{\pi^+}^{-1},
\tilde{G}_{\pi^-}^{-1}, \tilde{G}_{\sigma}^{-1})$, and $\tilde{B}^b$
is the transformed coupling matrix, which depends on $\omega_n$ as
was mentioned earlier. Similar expressions can be derived in case of
fermions.

\section{THE PARAMETERIZATION}
\label{param}

Before calculating at finite temperature and non--zero
chemical potentials one has to parameterize the model at
$T=\mu_{\mathrm{I,B}}=0$. We closely follow the method presented in
\cite{herpay06, kovacs06, kovacs07}. Since $\rho=0$ at $\muI=0$, there are five
parameters, namely $m^2$, $\lambda$, $g_F$, $h$ and $v$, which
can be fixed by setting four physical quantities ---namely the pion,
sigma, u and d quark masses plus the pion decay constant (through the PCAC
relation)--- to their physical values and by requiring the fulfillment
of the \eqref{e:ward} equation of state. At $T=\mu_{\mathrm{I,B}}=0$ and
$\rho=0$, the one--loop level $\pi_3$ inverse propagator can be written as
\be
\label{e:prop_1}
\im (G^{\textrm{1--loop}}_{\pi_3})^{-1}={(1-\delta Z)p^2 -m^2 -\delta m^2
  -(\lambda +\delta \lambda) v^2 -\Sigma_{\pi_3}(p^2=0)- p^2
  \Sigma_{\pi_3}^{\prime}(p^2=0) -\tilde{\Sigma}(p^2)},
\ee
where $\Sigma_{\pi_3}^{\prime} = \partial\Sigma_{\pi_3}/\partial p^2 $
and $\tilde{\Sigma}_{\pi_3} \sim {\cal O} (p^4)$. Fixing the physical
$\pi_3$ mass ($M_\pi=138$\,MeV) through the $\pi_3$ one--loop level
propagator at $p^2=0$ and using the \eqref{e:mass_eq} mass resummation
equation at $\rho=T=\mu_{\mathrm{I,B}}=0$ one obtains
\be
\label{e:pion_par}
M_\pi^2=m^2+\lambda v^2+3\lambda(T_0^b(M_\pi,l_b)+T_0^b(m_\sigma,l_b))+
2g_F^2N_c T^{f,\pi}_0(m_f,l_f),
\ee
where $m_\sigma^2=M_\pi^2+2\lambda v^2$ and $T_0^{b/f,\pi}$ denotes
the bosonic/fermionic tadpole integrals at zero $T$ and
$\mu_{\mathrm{I,B}}$. In \eqref{e:pion_par} the lack of bubble
integrals is due to the fact that at $p^2$=0 they reduce to a linear
combination of tadpoles. These tadpoles are finite on account of the
$\delta m^2$ and $\delta \lambda$ counterterms \eqref{e:cnt}. Note
that new divergent terms do not appear in the tadpoles at finite
temperature and/or non--zero chemical potentials, as it should be. 
The presence of $\delta Z$ in \eqref{e:prop_1}  is also important 
because this term
renders the $p^2$ dependent part of the propagator finite. It involves
also a finite renormalisation:
\be
\delta Z_{\pi_3}^{\mathrm{fin}}=\delta Z + \frac{\partial\Sigma_{
    \pi_3}}{\partial p^2}(p^2=0) = \frac{\lambda}{16\pi^2}
\frac{M_\pi^4 -m_\sigma^4 +M_\pi^2 m_\sigma^2 \log\left(
    \frac{m_\sigma^2} {M_\pi^2}\right)} {(m_\sigma^2-M_\pi^2)^2} +
\frac{g_F^2 N_c} {16\pi^2} \log\left(\frac{\mathrm{e} m_f^2}{l_f^2}\right)
\ee
and the one--loop level PCAC relation depends on $\delta Z_{\pi_3}^{
  \mathrm{fin}}$ as follows
\be
v [\im  G^{\textrm{1-loop}}_{\pi_3}(p^2=0)]^{-1} = f_\pi M_\pi^2 (1 -
\delta Z_{\pi_3}^{\mathrm{fin}}/2),
\ee
where $f_\pi=93$\,MeV is the pion decay constant. Since $\delta
Z_{\pi_3}^{ \mathrm{fin}}$ depends on the fermionic renormalization
scale its actual value is tunable. Thus we required the
vanishing of $\delta Z_{\pi_3}^{ \mathrm{fin}}$ to fix the fermionic
renormalization scale. This requirement makes the PCAC relation
simpler and thus the values of $v$, $g_F$, $h$ offer themselves
immediately
\be
v=f_\pi,\qquad g_F=2 \frac{m_f}{f_\pi},\qquad h=f_\pi M_\pi^2,
\ee   
where $m_f=938/3$\,MeV is the constituent $u,d$ quark mass. At
this point $v$, $g_F$, and $h$ are known and $m^2$ can be expressed from
\eqref{e:pion_par}.

The remaining unknown $\lambda$ parameter is
determined by fixing the physical $\sigma$ mass at one--loop level and
at zero external momentum, that is
\bea
M_\sigma^2& = &m^2 + \delta m+ 3(\lambda+\delta\lambda)v^2 +
\Sigma_\sigma(p^2=0)\no\\
& = & m^2 + 3\lambda v^2 + 3\lambda(T_0^b(M_\pi,l_b) + T_0^b(m_\sigma,
l_b)) + 18\lambda^2 v^2 B_0^b(m_\sigma, l_b) + 6\lambda^2 v^2
B_0^b(M_\pi, l_b) + 6 g_F^2 T_0^{f, \sigma}(m_f, l_f),
\eea
where $B_0^b$ and $T_0^{f, \sigma}$ come from the temperature and
chemical potential independent part of the bosonic/fermionic bubble
diagram at zero external momentum and degenerate masses. It is worth
to note that the same infinite counterterms
render the above equation  as well as the equation of $M_{\pi}$
finite.
\begin{figure}[!t]
\includegraphics[width=0.95\textwidth]{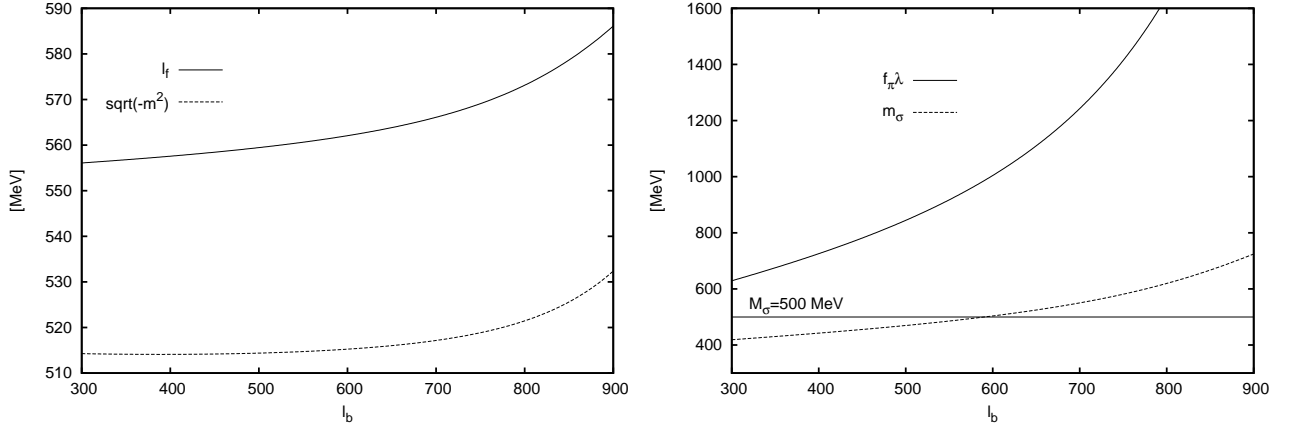}
{\caption{Bosonic renormalization scale dependence of the 
    fermion renormalization scale $l_f$, the mass 
    parameter $m^2$ (left panel) and of the coupling $\lambda$, the tree
    level $\sigma$ mass $m_{\sigma}$ (right
    panel).}\label{f:par}}
\end{figure}

We use $M_{\sigma}=500$\,MeV for the physical $\sigma$ mass. This choice
seems somewhat arbitrary, because the $\sigma$ meson is a broad
resonance rather than a particle with well--defined mass (see
\cite{prd06} and references therein). Thus it would be more
appropriate to identify the mass and width of the $\sigma$ meson
through the pole of its spectral function \cite{jakovac03}. However,
we checked that varying $M_\sigma$ in the 400\,MeV\,--\,750\,MeV range
produces just the same order of uncertainties in the thermodynamical
results as the variation of the $l_b$ boson renormalization scale itself.
Accordingly, it is enough to analyze the $l_b$ dependence of the
different parameters, hence here and in the following we present
our results which correspond to the choice $M_\sigma=500$\,MeV at
zero $T$, $\muIB$. On the left panel of Fig.~\ref{f:par} the
bosonic renormalization scale dependence of  $m^2$ and $l_f$, while
on the right panel the tree level $\sigma$ mass and the $\lambda$
parameter are shown. As one can see on the right panel, the tree level
$\sigma$ mass equals its one\,--\,loop level value at $l_b\approx
600$\,MeV thus  the $\sigma$ mass becomes selfconsistent at this point
(but only for zero $T$, $\muIB$!). In addition $m$ and $l_f$ moderately depend on the
renormalization scale around this point (see left panel of
Fig.~\ref{f:par}). According to the above arguments we choose
the following scale range: $l_b\in [400$\,MeV, $800$\,MeV$]$ for the
thermodynamical calculations.

\section{RESULTS AT LOWEST ORDER IN $\rho$}

As was derived in Sec. \ref{eqns_1_loop} the  expectation value $\rho$ of the
pion field is determined by \eqref{e:rho_pert} up to second order in
$\rho$. From this equation it can be
immediately seen that the second order boundary for the occurrence of
the pion condensation in the $\muI-\muB-T$ space is determined by
\be
\label{e:condition}
\mu_I^2 -m^2- \lambda v^2 - r^{(0)}(m^2_{\pi_3}, v, T,\muIB)=0.
\ee
Moreover, the $\muIB$ and  $T$ dependence of $v$ and $m_{\pi_3}$ are
determined by \eqref{e:ward} and \eqref{e:pi3_pert}, which have to be solved at $\rho=0$. At $\rho=0$
\eqref{e:pi3_pert} has the same form as  \eqref{e:pion_par} with the
slight difference that now the tadpoles which appear therein have to
be calculated
at finite temperature and chemical potentials, while the form of
\eqref{e:ward} is unchanged. 
Using the explicit expression of $r^{(0)}$, Eq.~\eqref{e:condition} can
be expressed as 
\be
\label{e:surface}
\mu_I^2 - m_{\pi_3}^2(T,\muI,\muB) - R^{\text{1--loop}}(T,\muI,\muB) =0,
\ee
where $R^{\text{1--loop}}$ is the remaining part of $r^{(0)}$ after
subtracting from it $m^2_{\pi_3}$. It contains one--loop bosonic and fermionic
contributions. From \eqref{e:surface} it is obvious that at one--loop level
the condensation does not start exactly at $\muI=m_{\pi_3}$, as it is
commonly expected, but it is shifted to some extent by
$R^{\text{1--loop}}$. It's worth to note that the deviation $R^{\text{1--loop}}$
does not vanish identically even if the resummed pion mass is defined
as the pole of the propagator.

\begin{figure}
\includegraphics[width=\textwidth]{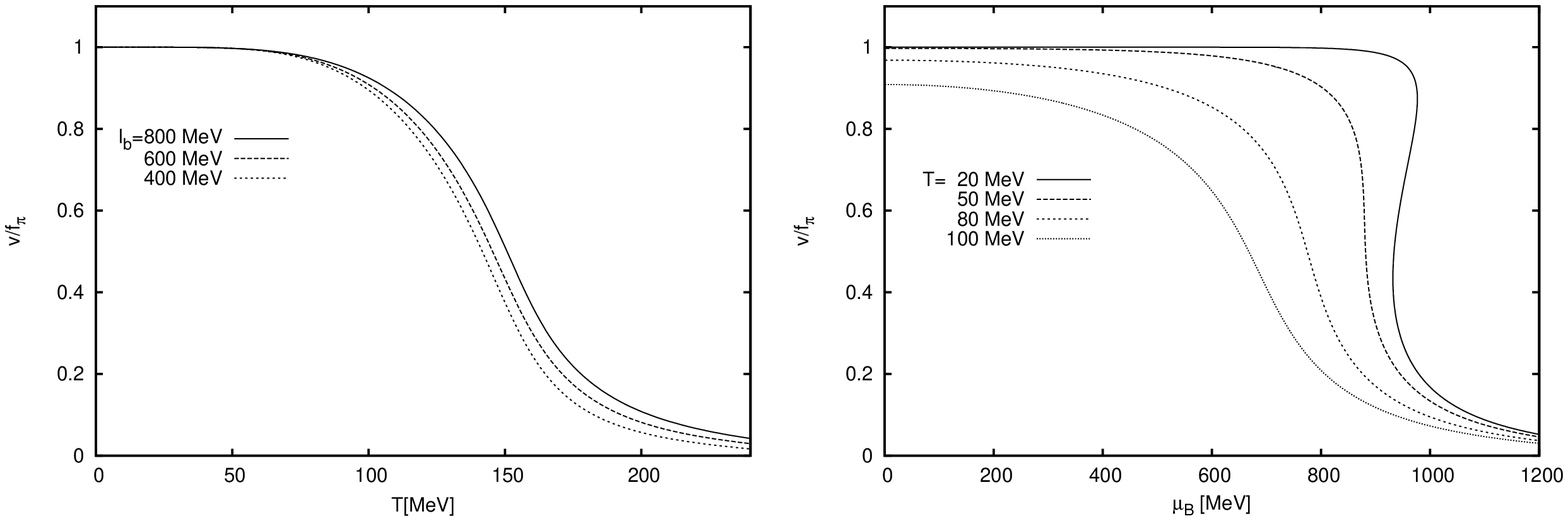}
{\caption{Temperature and chemical potential dependence of the 
    scalar condensate $v$. Left panel shows $v(T)$ at $\muIB=0$ and
    $l_b=400, 600, 800$\,MeV. Right panel shows $v(\muB)$ at $\muI=0$ and
    $T=20, 50, 80, 100$\,MeV.}\label{f:v}}
\end{figure}

First, we investigated the temperature and chemical potential
dependence of the scalar condensate $v$ at $\muI=0$ and different
values of $l_b$ by solving \eqref{e:ward} and \eqref{e:pi3_pert},
which can be seen in Fig.~\ref{f:v}. 
On the left panel the temperature dependence of $v$ is shown at
$l_b=400, 600, 800$\,MeV and at zero chemical potentials. As it is
expected, the
scalar condensate  shows a smooth crossover as the
chiral symmetry is restored at a pseudocritical temperature ($T_c$)
 around $150$\,MeV, which is in good agreement with the continuum
limit lattice result found in \cite{aoki06}. Moreover, the $v(T)$ curve
and consequently $T_c$ slightly depend on $l_b$. 
Hence, in the forthcoming we use
the fixed scale $l_b=600$\,MeV. On the right panel the baryochemical
potential dependence of $v$ can be seen at $l_b=600$\,MeV scale, and
at $T=20, 50, 80, 100$\,MeV temperatures. At small temperatures the
transition is of first order, while for large temperatures it
is of analytic crossover type, indicating the existence of a critical
endpoint (CEP), where the transition changes from first order to
crossover with increasing temperature (see e.g. in \cite{kovacs06} and
references therein). As one can see on Fig.~\ref{f:v} the temperature
at the CEP is around $50$\,MeV, which is much lower than the lattice
result presented in Ref. \cite{fodor04}, however this is a common feature
of effective models (for two/three flavors see
e.g. Refs.~\cite{jakovac03}, \cite{kovacs06}). The
critical/pseudocritical baryochemical potential values range from
$\sim600$\,MeV to $\sim1000$\,MeV depending on the
temperature.

\begin{figure}
\includegraphics[width=0.65\textwidth]{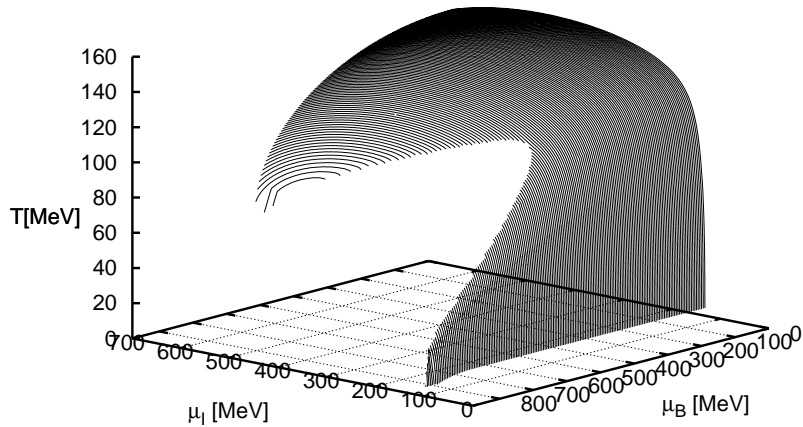}
{\caption{Second order critical surface of the pion condensation in
    the $\muI-\muB-T$}\label{f:surface}}
\end{figure}

Next, solving \eqref{e:ward} and \eqref{e:pi3_pert} for different $T, \muI,
\muB$ values and tracking the fulfillment of the \eqref{e:surface}
condition we determined the second order critical surface of
pion condensation,
which can be seen on Fig.~\ref{f:surface}.
As a function of the isospin chemical potential at fixed $\muB$
 the region of pion condensation starts very steeply at around
$130$\,MeV. After that steep jump a plateau starts, which decreases
slowly towards higher values of $\muI$. Moreover, it can be seen that
the region of pion condensation shrinks with increasing $\muB$ and it
even disappears at around $\muB=830$\,MeV (and $\muI=131$\,MeV), a
behavior which is in accordance with previous effective model studies (see
e.g. Sec.~VII.A of \cite{he05}). This is understandable physically,
since at large $\muB$ the condensate $\rho$ is basically determined by
the difference of the $u-d$ quark contributions in the EoS, and this
difference is decreasing with increasing $\muB$ at fixed $\muI$,
because the Fermi--Dirac factor in the contribution of $u$ depends on
$\muB+\muI$, while in the contribution of $d$ on $\muB-\muI$.
It is worth to observe that on the surface a gradually increasing missing
part starts from about $\muB=415$\,MeV and $\muI=221$\,MeV. In that
region the $\mu_I^2 -m^2 -\lambda v^2-r^{(0)}$
combination (numerator of Eq. \eqref{e:rho_pert}) is negative, which
means that the transition is not second order anymore. Strictly
speaking if in this region there is a nonzero solution to $\rho$ this
can happen only if $\lambda+r^{(2)}<0$ (denominator of
Eq. \eqref{e:rho_pert}) and in this case the transition is of first
order.

\begin{figure}
\includegraphics[width=0.5\textwidth]{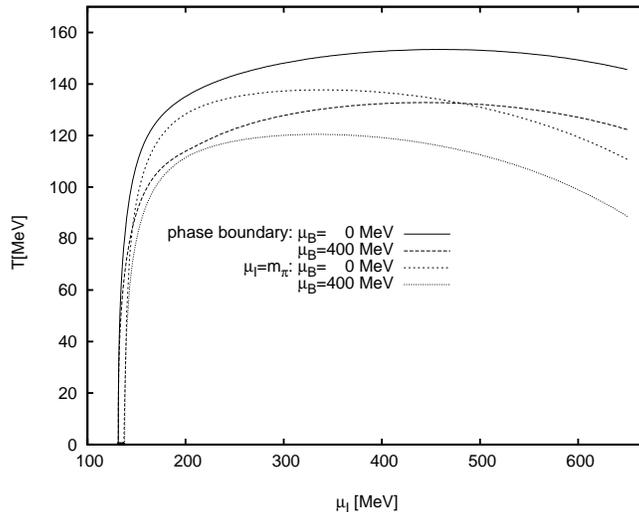}
{\caption{Phase boundary of pion condensation and the
    $\muI=m_{\pi_3}$ condition as function of the isospin chemical
    potential for different $\muB$ values.}\label{f:slice}}
\end{figure}

To analyze the surface in detail, two sections taken at $\muB=0$\,MeV and
$\muB=400$\,MeV are plotted in Fig. \ref{f:slice} together with the
$\muI=m_{\pi_3}$ curves.
At both baryochemical potentials the condensation starts at around
$\muI=131$\,MeV, which is slightly below the
$m_{\pi_3}=138$\,MeV pion mass, this deviation is due to the
corrections $R^{\text{1--loop}}$ in (\ref{e:surface}). Moreover, the
deviation widens as $\muI$ increases. 
Another interesting feature is that at fixed high temperature as we increase
the isospin chemical potential the condensation evaporates above a
certain $\muI$ value, which is in accordance with
Ref. \cite{zhang06}, where this phenomenon was observed in case of
two flavor PNJL model. In the region where the condensate already
evaporated ($\rho=0$) the chiral symmetry is almost totally restored
($v\approx 0$).

\begin{figure}
\includegraphics[width=0.5\textwidth]{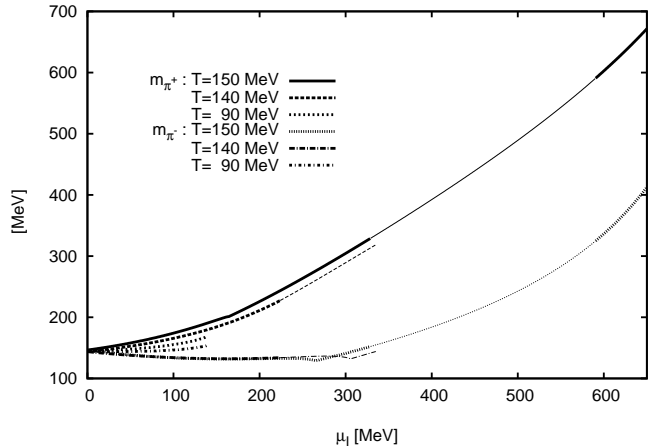}
{\caption{$\muI$ dependence of the one--loop pole masses of charged
    pions at different temperatures ($T=90,140,150$\,MeV). The thin lines
    indicate the condensed phase for each temperature (where our
    calculation is not valid).}\label{f:mass}}
\end{figure}

Finally we calculated the one--loop pole masses of
the charged pions on different temperatures as a function of the
isospin chemical potential at $\rho=0$. For this we calculated the
  self energies of the charged pions, then we solved the following
  pole mass equations:
  \be 
  \left(M_{\pi^{\pm}}^{\text{pole}}\right)^2= \left(m_{\pi^{\pm}}^{\text{tree}}\right)^2 +
  \Sigma_{\pi^{\pm}}( \omega = M_{\pi^{\pm}}^{\text{pole}}, \p = 0, T, \muIB),
  \ee
  where we have implicitly used the already known solutions
  $v(T,\muIB)$ and
  $m_{\pi_3}(T,\muIB)$. The one--loop masses are plotted on Fig. \ref{f:mass}.
On the figure thick lines represent the sectors, where our calculation
is valid, that is $\rho=0$. At the thin line parts of the curves the
pions are condensed, thus we should go beyond the lowest
approximation in $\rho$ to get correct results in that sectors. It is
worth to note that when the condensation sets in none of the one--loop
charged pion masses becomes zero as opposed to the masses defined
through the dispersion relation (see e.g. \cite{andersen06}). Another
interesting thing is that at larger values of $\muI$ where the
condensation has already evaporated and our calculation is valid (thick
line parts) the charged pion masses are still different, however
$\rho=0$. This
difference is due to the fact that the isospin background acts
differently on the charged pions (with opposite signs).

\section{CONCLUSIONS}
\label{sec:conclusion}

In this paper we studied the pion condensation in the framework of
$SU(2)_L\times SU(2)_R$ constituent quark model with explicitly broken
symmetry term  in the presence of baryochemical potential.
The model was parameterized at one--loop level and optimized perturbation
theory was used for the resummation of the perturbative series. The
one--loop equations were expanded in powers of $\rho$, and a Landau type
analysis was performed for the phase boundary at lowest order in
$\rho$. A simple condition for the boundary of the pion condensation
was set up, and we argued that this condition gives a second order
surface in the $\muI-\muB-T$ space. The temperature and
renormalization scale dependence of the scalar condensate $v$ was
investigated, and a mild  renormalization scale
dependence was found. At zero baryochemical potential the
pseudocritical temperature is in
accordance with results found on lattice \cite{aoki06}. Using the
condition for the pion condensation the second order surface was
determined. It was found that the surface starts steeply with
increasing $\muI$ at fixed $\muB$ and towards large values of $\muB$
the pion condensed region shrinks and even disappears at around
$\muB=830$\,MeV. However, at such a high energy one should take into
account the effects of the strange quark. Investigating different
sections of the surface it was showed that at one--loop level the pion
condensation curve slightly differ from the $\muI=m_{\pi_3}$ curve at
small $\muI$ and
this deviation increases with increasing $\muI$. We also studied the
dependence of the charged one--loop pion masses on
the isospin chemical potential. As a continuation of the present study
the analysis can be extended to higher order in $\rho$, with which for
instance the scaling properties around the surface, the dispersion
relation at one--loop level and different phases of the condensed
matter (BCS/LOFF) could be investigated.

\section*{Acknowledgment}

Work supported by the Hungarian Scientific Research Fund (OTKA) under
contract numbers NI68228 and T046129. We thank A. Patk{\'o}s and
Zs. Sz{\'e}p for suggestions and careful reading of the manuscript.

\appendix*

\section{Couplings}
\label{a:couplings}

The coupling matrices appearing in \eqref{e:sum2}, \eqref{e:eos2}
and \eqref{e:eosr} can be obtained from the interaction term of the
shifted fields in the Lagrangian 
\be
\frac{\lambda}{4}\phi^4
+\frac{g_F}{2}\bar{\psi}T_{\alpha}\phi_{\alpha}\psi\longrightarrow
\frac{\lambda}{4}\left((\pi_1+\rho)^2 + \pi_2^2 + \pi_3^2 +
(\sigma+v)^2\right)^2 + \frac{g_{F}}{2} \bar{\psi} \left(\tau_0
(\sigma+v) + \im \gamma_5(\tau_1(\pi_1+\rho) +\tau_2\pi_2 +\tau_3\pi_3)
\right)\psi\label{e:sh2}.
\ee
The four and three point couplings of \eqref{e:sh2}
determine the coefficients of the tadpole and bubble terms of
\eqref{e:sum2}. Including the symmetry factors of the corresponding
graphs these are:
\be
T^b=\begin{pmatrix} 1 & 0 & 0 & 0 \\ 0 & 1 & 0 & 0 \\  0 & 0 & 1 & 0
\\  0 & 0 & 0 & 3 \end{pmatrix}, \qquad B^b = \sqrt{2} \begin{pmatrix}
  0 & 0 & 0  & \rho \\ 0 & 0 & 0 & \rho \\ 0 &  0 &  0 & \sqrt{2}
  v\\  \rho  & \rho & \sqrt{2} v & 0  \end{pmatrix}, \qquad B^f
= \frac{\im}{2}\begin{pmatrix} -\gamma_5 & 0 \\ 0 &
  \gamma_5  \end{pmatrix}, 
\ee
where in case of bosons the same convention is used for the labeling
of the matrix elements as in \eqref{prop_b}. The bosonic bubble
contribution in \eqref{e:sum2} can be rewritten as
\be
\sum_{i,j,k,l} B^b_{ij}B^b_{kl}( G^b_{il}G^b_{jk} +G^b_{jl}G^b_{ik})=2
\Tr\{G^{b\,T} B^b G^b B^b\}, \label{e:tr_b}
\ee
and  by virtue of the structure of $B^b$ and $G^b$ one can factor out
the $\pi_3$ propagator from the above expression as follows
\be
\label{a:bubble_tr}   
 \Tr\{G^{b\,T} B^b G^b B^b\}=G^b_{\pi_3}\Tr \{{B^b}^{\prime}
 G^b\},
\ee
where  
\be
{B^b}^{\prime} = 2 \begin{pmatrix} \rho^2 & \rho^2 & \sqrt{2}v \rho & 0
  \\ \rho^2 & \rho^2 & \sqrt{2}v \rho & 0 \\ \sqrt{2} v\rho & \sqrt{2}
  v\rho & 2v^2  &  0\\   0  & 0 & 0 & 0 \end{pmatrix}.
\ee
The fermionic bubble contribution can be formulated as,
\be
\sum_{i,j,k,l} G^f_{il}B^f_{ij}G^f_{jk}B^f_{kl}=
\Tr\{G^{f\,T} B^f G^f B^f\}=\Tr\{B^f G^fB^fG^{f\,T}\},  \label{e:tr_f}
\ee
where the trace is over flavor as well as Dirac indices.

Moreover, in equations of states \eqref{e:eos2} and \eqref{e:eosr} 
the tadpole coefficients are also determined by the three point
couplings of \eqref{e:sh2},  
\be
R^b=\begin{pmatrix} 2\rho & \rho & v/\sqrt{2} & 0 \\  \rho & 2\rho &
v/\sqrt{2} & 0 \\  v/\sqrt{2} &   v/\sqrt{2} & \rho & 0 \\ 0 & 0& 0&
\rho \end{pmatrix},\qquad R^f=\frac{1}{2}\begin{pmatrix} 1 & 0 \\ 0 &
1 \end{pmatrix}, 
\ee
and
\be
H^b=\begin{pmatrix} v & 0 &\rho/\sqrt{2} & 0 \\ 0 & v & \rho/\sqrt{2}
&0 \\ \rho/\sqrt{2}&  \rho/\sqrt{2} & 3 v & 0\\ 0 & 0 & 0 &
v \end{pmatrix}, \qquad H^f=\frac{\im}{2}\begin{pmatrix} 0 & \gamma_5
  \\ \gamma_5 & 0 \end{pmatrix}.
\ee

\end{document}